\documentclass[useAMS,usenatbib]{mn2e}
\usepackage{graphicx}
\usepackage{float}

\floatstyle{ruled}
\newfloat{algorithm}{tbp}{loa}
\providecommand{\algorithmname}{Algorithm}
\floatname{algorithm}{\protect\algorithmname}

\title[MUSIC for Faraday Rotation Measure Synthesis]{MUSIC for Faraday Rotation Measure Synthesis}

\author[M. Andrecut]{M. Andrecut\thanks{E-mail:
mandrecu@ucalgary.ca; mircea.andrecut@gmail.com} \\
Institute for Space Imaging Science, University of Calgary, Calgary, Alberta, T2N 1N4, Canada}

\begin{document}

\date{Accepted 1988 December 15. Received 1988 December 14; in original form 2012 October 15}

\pagerange{\pageref{firstpage}--\pageref{lastpage}} \pubyear{2002}

\maketitle

\label{firstpage}

\begin{abstract}
Faraday Rotation Measure (RM) synthesis requires the recovery of the
Faraday Dispersion Function (FDF) from measurements restricted to
limited wavelength ranges, which is an ill-conditioned deconvolution
problem. Here, we propose a novel deconvolution method based on an
extension of the MUltiple SIgnal Classification (MUSIC) algorithm. The complexity and speed of the method is determined by the eigen-decomposition of the covariance matrix of the observed polarizations. 
We show numerically that for high to moderate Signal to Noise (S/N) cases 
the RM-MUSIC method is able to recover the Faraday depth values of
closely spaced pairs of thin RM components, even in situations where
the peak response of the FDF is outside of the RM range between the
two input RM components. This result is particularly important because
the standard deconvolution approach based on RM-CLEAN fails systematically
in such situations, due to its greedy mechanism used to extract the RM components. For low S/N situations, both the RM-MUSIC and RM-CLEAN methods provide similar results. 
\end{abstract}


\section{Introduction}

Faraday rotation is a physical phenomenon where the position angle
of linearly polarized radiation propagating through a magneto-ionic
medium is rotated as a function of frequency. Faraday Rotation Measure
(RM) synthesis is an important method for analyzing multichannel polarized
radio data, where multiple emitting regions are present along the
single line of sight of the observations \citep{b1}. In practice, the method
requires the recovery of the Faraday Dispersion Function (FDF) from 
measurements restricted to limited wavelength ranges, which is an 
ill-conditioned deconvolution problem, raising important computational 
difficulties. At least four different approaches have been proposed so far  
to solve this problem. A first approach uses an adaptation of the CLEAN  algorithm \citep{h2} to the RM deconvolution (RM-CLEAN) \citep{h1}.
The second approach is wavelet-based, and assumes field symmetries
in order to project the observed data onto $\lambda^{2}<0$ \citep{f2}.
The third approach relies on nonlinear model fitting \citep{f1},
while the fourth approach is based on the novel compressed sensing (CS) paradigm \citep{l1,a2}. Whether these methods are successful or not in detecting the 
structure of the FDF depends on the Signal to Noise ratio (S/N), the separation 
of the RM components, the relative angle of the RM components at the 
observed wavelengths, and the wavelength range. For example, the current existing methods fail to recover the correct RM components when the separation
among them is below the resolution determined from the half maximum
of the main peak of the RM spread function (RMSF). Also, another major
problem is related to situations where the interference of pairs of
RM components conspire to place the peak response of the FDF outside
of the RM range between the two input RM components. In this case, the 
standard RM-CLEAN method fails, due to its intrinsic greedy mechanism used to pick up the RM components. 
Here we discuss a novel approach, which addresses these two 
significant problems from a different perspective. 
The proposed method is an extension of the MUltiple SIgnal Classification 
(MUSIC) algorithm, which is based on the eigen-decomposition (ED)
of the covariance matrix of the observed data \citep{c1}. The complexity and 
the speed of the method is therefore determined by the size of the 
eigen-decomposition problem, which for several hundreds of data points 
is comparable to the RM-CLEAN approach. Our numerical results show that 
for high to moderate S/N cases, the RM-MUSIC method gives very good results, outperforming the standard approach based on RM-CLEAN. 
For low S/N situations, both the RM-MUSIC and RM-CLEAN methods provide similar results. We should note that contrary
to the other existing methods, the RM-MUSIC method recovers only the
Faraday depth values. Once the Faraday depth values are determined,
the real and imaginary parts of the RM components can be easily computed
using the linear least squares fitting approach.

\section{RM-Synthesis}

The Faraday depth (in $\mathrm{rad}\,\mathrm{m}^{-2}$) is defined
as:
\begin{equation}
\phi(r)=0.81\int_{source}^{observer}n_{e}\mathbf{B}\cdot \mathbf{dr},
\end{equation}
 where $n_{e}$ is the electron density (in $cm^{-3}$) , $B$ is
the magnetic field (in $\mu G$), and $dr$ is the infinitesimal path
length (in parsecs). We also define the complex polarization as:
\begin{equation}
P(\lambda^{2})=Q(\lambda^{2})+iU(\lambda^{2})=pIe^{2i\chi(\lambda^{2})},
\end{equation}
where $p$ is the fractional polarization, $I$, $Q$, $U$ are the
observed Stokes parameters, and $\chi(\lambda^{2})$ is the polarization
angle observed at wavelength $\lambda$. Also, we assume that the
observed polarization $P(\lambda^{2})$ originates from the emission
at all possible values of $\phi$, corresponding to the Fourier transform:

\begin{equation}
P(\lambda^{2})=\int_{-\infty}^{+\infty}F(\phi)e^{2i\phi\lambda^{2}}d\phi,
\end{equation}
 where $F(\phi)$ is the complex FDF (the intrinsic polarized flux,
as a function of the Faraday depth). Thus, in principle $F(\phi)$
is the inverse Fourier transform of the observed quantity $P(\lambda^{2})$:
\begin{equation}
F(\phi)=\int_{-\infty}^{+\infty}P(\lambda^{2})e^{-2i\phi\lambda^{2}}d\lambda^{2}.
\end{equation}
 However, this operation is ill-defined since we cannot observe $P(\lambda^{2})$
for $\lambda^{2}<0$, and also in general the observations are limited
to an interval $[\lambda_{min}^{2},\lambda_{max}^{2}]$.

In order to deal with the above limitations, the observed polarization
is defined as:
\begin{equation}
\tilde{P}(\lambda^{2})=W(\lambda^{2})P(\lambda^{2}),
\end{equation}
 where $W$ is the observation window function, with $W(\lambda^{2})>0$
for $\lambda^{2}\in[\lambda_{min}^{2},\lambda_{max}^{2}]$, and $W(\lambda^{2})=0$
otherwise. Therefore, for $N$ measurement channels with frequencies
$\nu_{n}$, wavelengths $\lambda_{n}=c/\nu_{n}$, $n=1,2,...,N$ ($c$
is the speed of light), weights $W(\lambda_{n}^{2})=W_{n}$, polarizations
$P(\lambda_{n}^{2})=P_{n}$, we obtain the following discrete expression for the FDF:
\begin{equation}
F(\phi)=A^{-1}\sum_{n=0}^{N-1}W_{n}P_{n}e^{-2i\phi(\lambda_{n}^{2}-\lambda_{r}^{2})},
\end{equation}
where 
\begin{equation}
A=\left[\sum_{n=0}^{N-1}W_{n}\right]^{-1}
\end{equation}
is a normalization constant, and the reference wavelength $\lambda_{r}^{2}$ is defined as:
\begin{equation}
\lambda_{r}^{2}=A^{-1}\sum_{n=0}^{N-1}W_{n}\lambda_{n}^{2}.
\end{equation}

In our approach we assume that the model of $F(\phi)$ contains $K\ll N$
(unknown) components $f_{k}\delta(\phi-\phi_{k})$, where $f_{k}$ are complex and 
$\phi_{k}$ are real quantities $k=0,...,K-1$:
\begin{equation}
F(\phi)=\sum_{k=0}^{K-1}f_{k}\delta(\phi-\phi_{k}).
\end{equation}
Considering additive noise in the measurement process, the sampled
polarization-domain channel response is given by:
\begin{equation}
P_{n}=W_{n}\sum_{k=1}^{K}f_{k}e^{2i\phi_{k}(\lambda_{n}^{2}-\lambda_{r}^{2})}+\alpha_{n}+i\beta_{n},\quad n=1,2,...,N,
\end{equation}
where $N$ is the number of measured channels. Here, both $\alpha_{n}$
and $\beta_{n}$ are Gaussian variables with mean zero and standard
deviation $\sigma$. Obviously, $F(\phi)$
is a ``dirty'' reconstruction, and a deconvolution step is necessary
to recover the RM components $f_{k}\delta(\phi-\phi_{k})$, given
the observed values $P_{n}$ and the weights $W_{n}$. 

\section{RM-MUSIC}

The MUSIC method is generally used for estimating frequencies in signal
processing problems, and the location of pointlike scatterers in imaging
problems (see \cite{c1} for a review). The standard MUSIC method is based on the ED of an Hermitian
operator, which corresponds to the covariance matrix of the signal
or the array response matrix. By the finite-dimensional spectral theorem,
such operators can be associated with an orthonormal basis of the
underlying space in which the operator is represented as a diagonal
matrix with real number entries. The main idea is to estimate the
frequencies, or to localize multiple sources, by exploiting the eigen-structure
of this Hermitian operator. More exactly the space spanned by its
eigenvectors can be partitioned into two orthogonal subspaces, namely
the signal subspace and the noise subspace. By exploiting the orthogonality
of the signal and noise subspaces, the MUSIC method significantly
improves the resolution (i.e. locating closely spaced frequencies
or scatterers), and as a consequence it is considered a super-resolution
method. Let us now formulate the standard MUSIC approach to the Faraday
depth recovery problem. 

Since only one snapshot of measurement data $P$ of length $N$ is
available, the data sequence is divided into $M=N-L$ segments of
length $L$, $0<L\leq N$, and then the $L\times L$ covariance matrix
is estimated as: 
\begin{equation}
\mathbf{\hat{G}}=\frac{1}{M}\sum_{m=0}^{M-1}\mathbf{P}^{(m)}(\mathbf{P}^{(m)})^{H}
\end{equation}
where $\mathbf{P}^{(m)}=[P_{m},P_{m+1},...,P_{m+L}]^{T}$, $m=0,1,...,M-1$,
and the superscript $H$ denotes the conjugate transpose operation. Suppose
that $\mu_{0}\geq\mu{}_{1}\geq...\geq\mu{}_{L-1}$ and $\mathbf{g}^{(0)},\mathbf{g}^{(1)},...,\mathbf{g}^{(L-1)}$
are the eivenvalues, and respectively the eigenvectors of $\mathbf{\hat{G}}$,
such that:
\begin{equation}
\mathbf{\hat{G}}\mathbf{g}^{(\ell)}=\mu_{\ell}\mathbf{g}^{(\ell)},
\end{equation}
and 
\begin{equation}
\mathbf{\hat{G}}=\sum_{\ell=0}^{L-1}\mu_{\ell}\mathbf{g}^{(\ell)}(\mathbf{g}^{(\ell)})^{H}.
\end{equation}
Since the measured signal contains only $K$ components, the last
$L-K$ eigenvalues of $\mathbf{\hat{G}}$, $\mu_{K}\geq\mu_{K+1}\geq...\geq\mu_{L-1}$,
should be small and below the noise level, and we say that the corresponding
eigenvectors $\mathbf{g}^{(K)},\mathbf{g}^{(K+1)},...,\mathbf{g}^{(L-1)}$
span the noise subspace of $\mathbf{\hat{G}}$, while the first $K$
eigenvectors $\mathbf{g}^{(0)},\mathbf{g}^{(1)},...,\mathbf{g}^{(K-1)}$
span the signal subspace, and the corresponding eigenvalues and eigenvalues
$\mu_{0}\geq\mu{}_{1}\geq...\geq\mu{}_{K-1}$ should be above the
noise level. Thus, the eigen-decomposition of $\mathbf{\hat{G}}$
can be written as: 
\begin{equation}
\mathbf{\hat{G}}=\underbrace{\sum_{\ell=0}^{K-1}\mu_{\ell}\mathbf{g}^{(\ell)}(\mathbf{g}^{(\ell)})^{H}}_{signal\, subspace}+\underbrace{\sum_{\ell=K}^{L-1}\mu_{\ell}\mathbf{g}^{(\ell)}(\mathbf{g}^{(\ell)})^{H}}_{noise\, subspace}.
\end{equation}
We can now form the projection operator onto the noise subspace defined as:
\begin{equation}
\mathbf{\hat{G}}_{\mathbf{g}}^{(noise)}=\sum_{\ell=K}^{L-1}\mathbf{g}^{(\ell)}(\mathbf{g}^{(\ell)})^{H},
\end{equation}
and consider the signal subspace sampling vector 
\begin{equation}
\mathbf{h}(\tilde{\phi})=[e^{2i\tilde{\phi}\tilde{\lambda}{}_{0}^{2}},e^{2i\tilde{\phi}\tilde{\lambda}{}_{1}^{2}},...,e^{2i\tilde{\phi}\tilde{\lambda}{}_{L-1}^{2}}]^{T},
\end{equation}
where the wavelengths are interpolated as following: 
\begin{equation}
\tilde{\lambda}{}_{\ell}^{2}=\left[\frac{c}{\nu_{N}-\frac{\ell}{L-1}(\nu_{N}-\nu_{1})}\right]^{2},\quad\ell=0,1,...,L-1,
\end{equation}
such that they cover the initial observation interval $[\lambda_{min},\lambda_{max}]$
We should note that since $L\leq N$, the Faraday depth is also linearly interpolated as  $\tilde{\phi}=\phi L/N$, and therefore we obtain:
\begin{equation}
\mathbf{h}(\phi)=[e^{2i\frac{L}{N}\phi\tilde{\lambda}{}_{0}^{2}},e^{2i\frac{L}{N}\phi\tilde{\lambda}{}_{1}^{2}},...,e^{2i\frac{L}{N}\phi\tilde{\lambda}{}_{L-1}^{2}}]^{T}.
\end{equation}
If the vector $\mathbf{h}(\phi)$ represents the signal, i.e. it is
a linear combination of the signal subspace eigenvectors, then its
projection onto the noise subspace must be close to zero:
\begin{equation}
\left\Vert \mathbf{\hat{G}}_{\mathbf{g}}^{(noise)}\mathbf{h}(\phi)\right\Vert \simeq0.
\end{equation}
Thus, the Faraday depth values $\phi_{k}$, $k=0,1,...,K-1$, should
correspond to the maxima of the following MUSIC pseudo-spectrum:
\begin{equation}
S(\phi)=\frac{1}{\left\Vert \mathbf{\hat{G}}_{\mathbf{g}}^{(noise)}\mathbf{h}(\phi)\right\Vert ^{2}}=\frac{1}{\sum_{\ell=K}^{L-1}\left\Vert (\mathbf{g}^{(\ell)})^{H}\mathbf{h}(\phi)\right\Vert ^{2}}.
\end{equation}

Once the Faraday depth values $\phi_{k}$ are determined from the
MUSIC pseudo-spectrum the real and imaginary parts of the RM
components $f_{k}$ can be easily computed using the linear least squares
fitting approach, i.e. by solving the following minimization problem:
\[
\{f_{k}\}_{k=0}^{K-1}=\arg\min_{f_{k}}\sum_{n=0}^{N-1}\left[P_{n}-\sum_{k=1}^{K}f_{k}e^{2i\phi_{k}\lambda_{n}^{2}}\right]^{2}
\]
\begin{equation}
=\arg\min_{\mathbf{f}}\left\Vert \mathbf{P} - \hat{\mathbf{\mathbf{\Psi}}}\mathbf{f} \right\Vert.
\end{equation}
This least-squares fit problem has fewer free parameters, and it is far better constrained than the fit that would have been done without MUSIC. In fact the problem has an unique minimum norm solution, obtained by solving the linear system of equations: 
$\mathbf{\hat{\mathbf{\Psi}}}\mathbf{f}=\mathbf{P}$, where $\mathbf{f}=[f_{0},f_{1},...,f_{K-1}]^{T}$
are the unknown components, $\mathbf{P}=[P_{0},P_{1},...,P_{N-1}]^{T}$
are the observed polarizations, and $\hat{\mathbf{\mathbf{\Psi}}}$ is
the Fourier matrix with the elements $\Psi_{n,k}=e^{2i\phi_{k}\lambda_{n}^{2}}$.
Thus, the least squares solution is given by $\mathbf{f}=\hat{\mathbf{\mathbf{\Psi}}}^{\dagger}\mathbf{P}$,
where $\hat{\mathbf{\mathbf{\Psi}}}^{\dagger}=(\hat{\mathbf{\mathbf{\Psi}}}^{H}\hat{\mathbf{\mathbf{\Psi}}})^{-1}\hat{\mathbf{\mathbf{\Psi}}}^{H}$
is the Moore-Penrose pseudo-inverse of $\hat{\mathbf{\mathbf{\Psi}}}$. 

\section{Numerical Implementation}

In order to optimally use the MUSIC method we also need to determine
the number of components $K$ in the polarization signal. In principle,
$K$ can be determined from the eigenvalues of the covariance matrix. 
The first largest eigenvalues,
which are much bigger than the noise level $\mu_{\ell}>\sigma^{2}$,
$\ell=0,1,...K-1$, should indicate the value of $K$. However, in
a practical implementation, when the covariance matrix is estimated
from a small number of observations, it is challenging to clearly
separate the signal eigenvalues from the noise eigenvalues. In these
ambiguous cases one can use information theoretic criteria for selection,
like the Minimum Descriptive Length (MDL) criteria, where the value
of $K$ corresponds to the minimum of the following quantity:
\[
H(K)=-M(M-K)\log\left(\frac{\prod_{m=K}^{M}\mu_{m}^{1/(M-K)}}{\frac{1}{M-K}\sum_{m=K}^{M}\mu_{m}}\right)\]
\begin{equation}
+\frac{1}{2}K(2L-K)\log M.\end{equation}
Also, since the smallest eigenvalues are not equal among them, one
can use the following normalized versions of the MUSIC pseudo-spectrum:
\begin{equation}
S(\phi)=\frac{1}{\sum_{\ell=K}^{L-1}\mu_{\ell}^{-1}\left\Vert (\mathbf{g}^{(\ell)})^{H}\mathbf{h}(\phi)\right\Vert ^{2}},
\end{equation}
that accounts for the variation of the eigenvalues, and it is less sensitive
to the $K$ estimation errors. Another influencing
parameter is the length of the data segments $L$. In order to preserve
as much information as possible in each data segment we should have $L\geq N/2$. 
Our numerical experiments have shown that $L=2N/3$ has
a good detection sensitivity, and this is the value used in the simulations
presented here. We should note also that the speed of the RM-MUSIC
depends on the parameter $L$, which gives the size $L\times L$ of
the eigen-decomposition problem. For $L$ in the order of hundreds,
the speed of RM-MUSIC is comparable to the speed of RM-CLEAN, or even
faster. The pseudo-code of the RM-MUSIC method is given in Algorithm 1.

\begin{algorithm}
\caption{Pseudo code of the RM-MUSIC method.}

$\mathbf{P}$; polarization data

$N$; number of channels

$N/2\leq L\leq N$; length of data segments

$M=N-L$; number of data segments

$\mathbf{\hat{G}}\leftarrow\frac{1}{M}\sum_{m=0}^{M-1}\mathbf{P}^{(m)}(\mathbf{P}^{(m)})^{H}$;
covariance matrix

$\{\mu_{\ell},\mathbf{g}^{(\ell)}\}_{\ell=0}^{L}\leftarrow ED(\mathbf{\hat{G}})$;
eigen-decomposition

$\{\tilde{\lambda}{}_{\ell}^{2}\}_{\ell=0}^{L}\leftarrow\left[\frac{c}{\nu_{N}-\frac{\ell}{L-1}(\nu_{N}-\nu_{1})}\right]^{2}$;
interpolation wavelengths

$\mathbf{h}(\phi)\leftarrow[e^{2i\frac{L}{N}\phi\tilde{\lambda}{}_{0}^{2}},e^{2i\frac{L}{N}\phi\tilde{\lambda}{}_{1}^{2}},...,e^{2i\frac{L}{N}\phi\tilde{\lambda}{}_{L-1}^{2}}]^{T}$;
sampling vector

$K\leftarrow\arg\min_{K}H(K)$; number of components 

$S(\phi)\leftarrow\frac{1}{\sum_{\ell=K}^{L-1}\mu_{\ell}^{-1}\left\Vert (\mathbf{g}^{(\ell)})^{H}\mathbf{h}(\phi)\right\Vert ^{2}}$;
MUSIC pseudo-spectrum

$\{\phi_{k}\}_{k=0}^{K-1}=\{\arg\max_{\phi}S(\phi)\}_{k=0}^{K-1}$; Faraday depths

$
\{f_{k}\}_{k=0}^{K-1}=
\arg\min_{\mathbf{f}}\left\Vert \mathbf{P} - \hat{\mathbf{\mathbf{\Psi}}}\mathbf{f} \right\Vert
$; complex amplitudes

return $\{\phi_{k},f_{k}\}_{k=0}^{K-1}$;
\end{algorithm}

\section{Numerical Results}

To illustrate numerically the RM-MUSIC method we consider the following
experiment layout: frequency range: $\nu_{min}=1100\,\mathrm{MHz}$,
$\nu_{max}=1400\,\mathrm{MHz}$; wave length range $\lambda_{min}^{2}=0.046\,\mathrm{m}^{2}$,
$\lambda_{max}^{2}=0.074\,\mathrm{m}^{2}$; number of observation
channels $N=150$; width of an observing channel $\triangle\nu=2\,\mathrm{MHz}$;
resolution in Faraday depth space $\delta\phi=2\sqrt{3}/(\lambda_{max}^{2}-\lambda_{min}^{2})\simeq122\,\mathrm{rad}\,\mathrm{m}^{-2}$;
equal weights $W_{n}=1$; noise level $\sigma=\sqrt{N}=12.247$. 

\begin{figure}
\centering
 \includegraphics[scale=0.855]{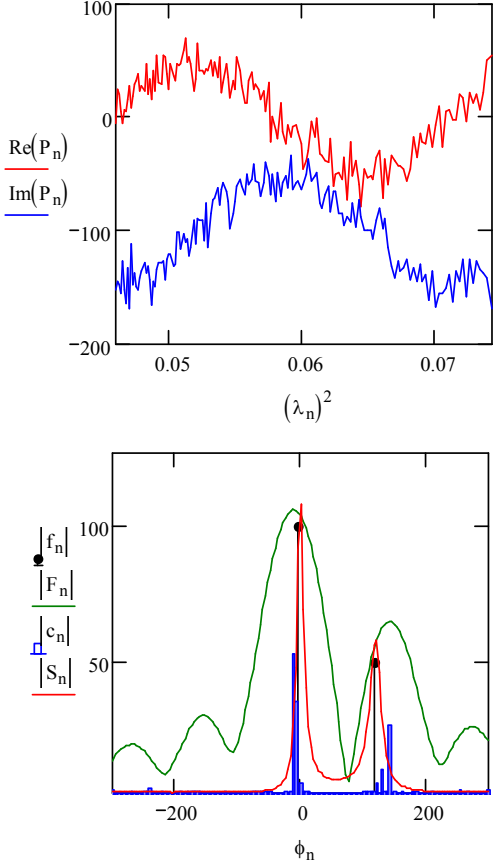}
 \caption{
RM-MUSIC successful result for two components with the complex amplitudes
$f_{1}=-100i$, $f_{1}=50i$, and Faraday depths $\phi_{0}=0\,\mathrm{rad}\,\mathrm{m}^{-2}$, $\phi_{1}=122\,\mathrm{rad}\,\mathrm{m}^{-2}$, separated by $\phi_{1}-\phi_{2}\simeq\delta\phi$ ($|f|$ component amplitudes, $|F|$  ``dirty'' FDF, $|c|$ RM-CLEAN components, $S$ RM-MUSIC spectrum). 
}
\end{figure}

\begin{figure}
\centering
 \includegraphics[scale=0.855]{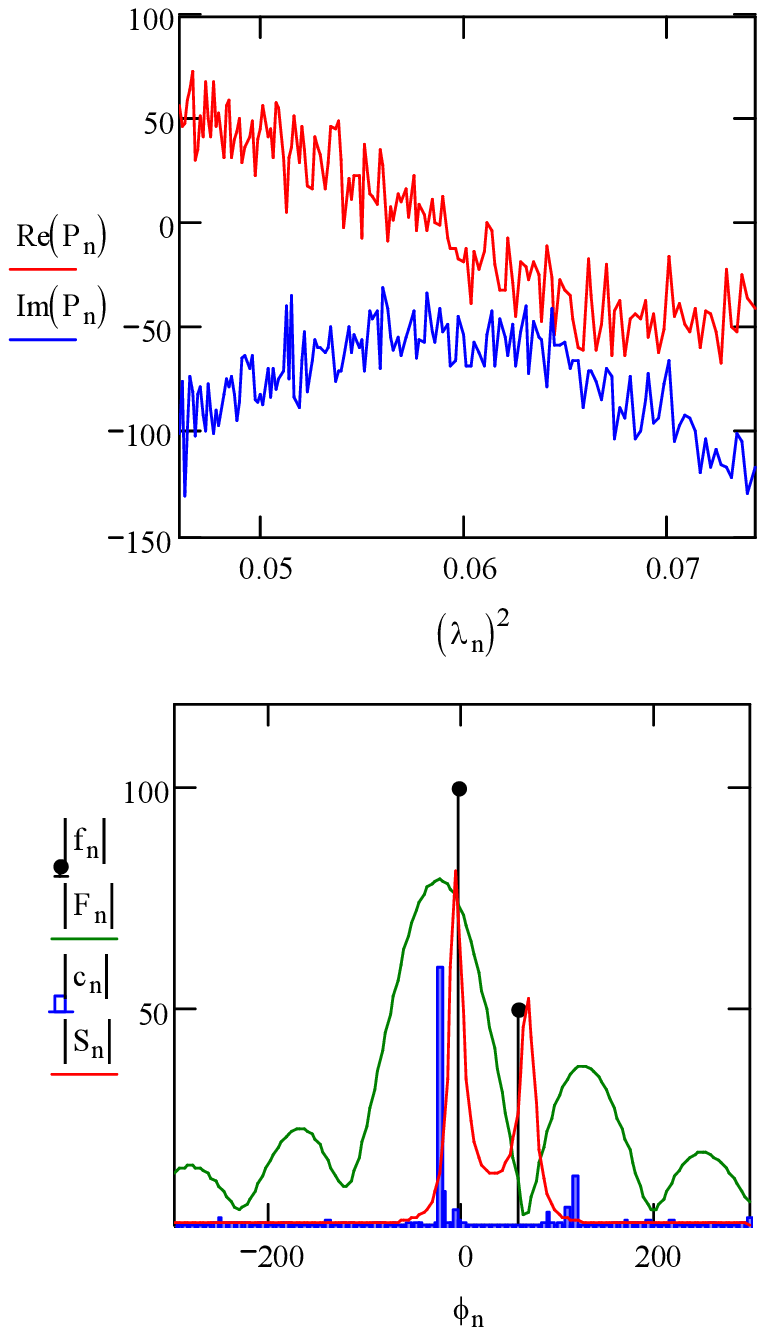}
 \caption{
RM-MUSIC successful result for two components with the complex amplitudes
$f_{1}=-100i$, $f_{1}=50i$, and Faraday depths $\phi_{0}=0\,\mathrm{rad}\,\mathrm{m}^{-2}$, $\phi_{1}=61\,\mathrm{rad}\,\mathrm{m}^{-2}$, separated by $\phi_{1}-\phi_{2}\simeq0.5\delta\phi$ ($|f|$ component amplitudes, $|F|$ ``dirty'' FDF, $|c|$ RM-CLEAN components, $S$ RM-MUSIC spectrum). 
}
\end{figure}

\begin{figure}
\centering
 \includegraphics[scale=0.855]{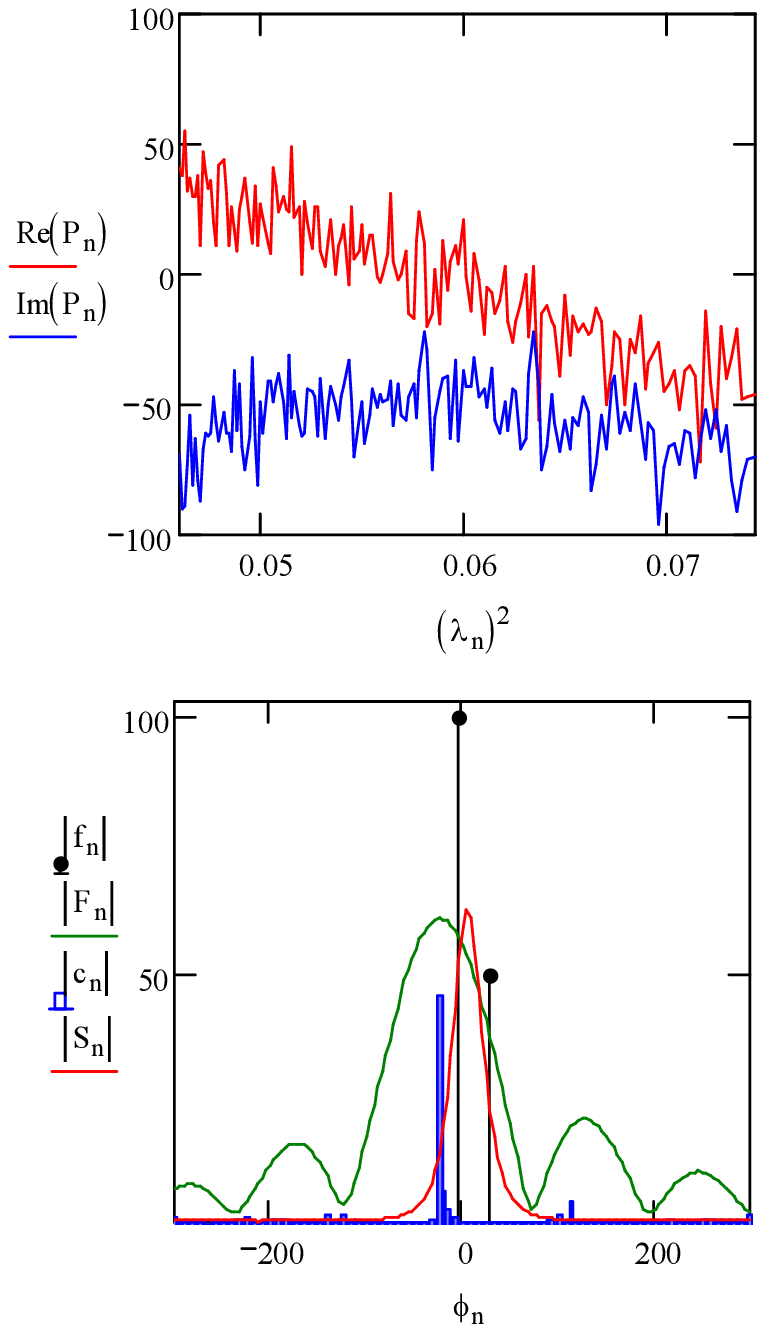}
 \caption{
RM-MUSIC result for two components with the complex amplitudes
$f_{1}=-100i$, $f_{1}=50i$, and Faraday depths $\phi_{0}=0\,\mathrm{rad}\,\mathrm{m}^{-2}$, $\phi_{1}=30.5\,\mathrm{rad}\,\mathrm{m}^{-2}$, separated by $\phi_{1}-\phi_{2}\simeq0.25\delta\phi$ ($|f|$ component amplitudes, $|F|$ ``dirty'' FDF, $|c|$ RM-CLEAN components, $S$ RM-MUSIC spectrum). 
}
\end{figure}

\begin{figure}
\centering
 \includegraphics[scale=0.855]{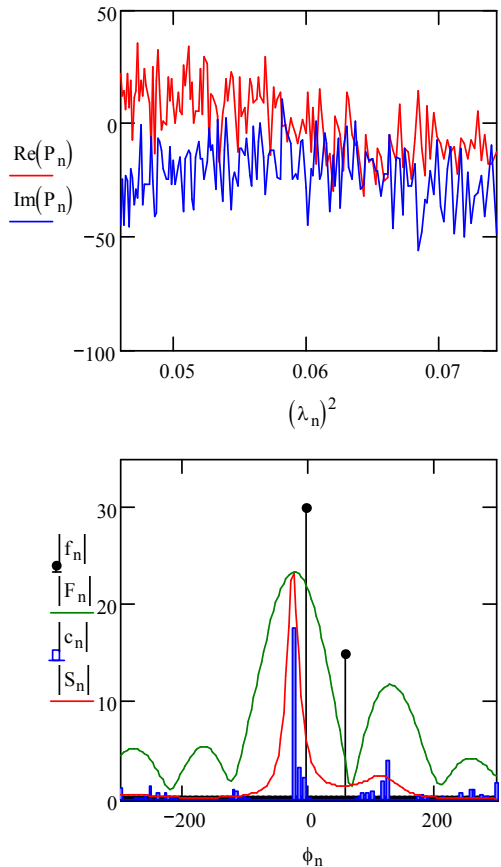}
 \caption{
RM-MUSIC result for low signal to noise is comparable to RM-CLEAN. Two components with 
$f_{1}=-100i$, $f_{1}=50i$, Faraday depths $\phi_{0}=0\,\mathrm{rad}\,\mathrm{m}^{-2}$, $\phi_{1}=61\,\mathrm{rad}\,\mathrm{m}^{-2}$, separated by $\phi_{1}-\phi_{2}\simeq0.5\delta\phi$ ($|f|$ component amplitudes, $|F|$ ``dirty'' FDF, $|c|$ RM-CLEAN components, $S$ RM-MUSIC spectrum). 
}
\end{figure}

The RM-MUSIC method works very well when the components are 
separated above the resolution limit $\delta\phi$. For example, in Figure 1 we have two components with the complex amplitudes $f_{0}=-100i$
and $f_{1}=50i$ at $\lambda^{2}=\lambda_{r}^{2}$, and Faraday depths
$\phi_{0}=0\,\mathrm{rad}\,\mathrm{m}^{-2}$, $\phi_{1}=122\,\mathrm{rad}\,\mathrm{m}^{-2}$,
separated by $\phi_{1}-\phi_{0}=\delta\phi$. For comparison
we also give the ``dirty'' FDF and the RM-CLEAN components. In order
to make the results comparable, we scaled $S(\phi)$ as $S(\phi)\leftarrow S(\phi)\left|F\right|_{max}\left|S\right|_{max}^{-1}$,
where $\left|F\right|_{max}$ and $\left|S\right|_{max}$ are the
maximum amplitudes of the original spectra. This way, both $S(\phi)$
and $\left|F(\phi)\right|$ are in the same range. One can see that
the RM-MUSIC spectrum recovers almost exactly the Faraday depths of
the two components $\tilde{\phi}_{0}=1.05\,\mathrm{rad}\,\mathrm{m}^{-2}$,
$\tilde{\phi}_{1}=122.81\,\mathrm{rad}\,\mathrm{m}^{-2}$, while both
the ``dirty'' FDF and RM-CLEAN fail to recover the Faraday depth
of the second component: $\tilde{\phi}_{0}=2.01\,\mathrm{rad}\,\mathrm{m}^{-2}$,
$\tilde{\phi}_{1}=142.95\,\mathrm{rad}\,\mathrm{m}^{-2}$. Moreover,
once the depths are recovered, the real and imaginary parts of the
RM components can be easily computed using a linear least squares
fitting approach, resulting in an almost exact recovery: $\tilde{f}_{0}=-1.95-99.74i$,
$\tilde{f}_{1}=0.05+50.02i$.

In Figure 2 we show that the MUSIC super-resolution method can give
good results even for separations below the the resolution in Faraday
depth space $\delta\phi$. Here, we have the same two components with
Faraday depths $\phi_{0}=0\,\mathrm{rad}\,\mathrm{m}^{-2}$, $\phi_{1}=61\,\mathrm{rad}\,\mathrm{m}^{-2}$,
separated by $\phi_{1}-\phi_{0}=0.5\delta\phi$. One can see
that both components are correctly separated by the RM-MUSIC spectrum
$\tilde{\phi}_{0}=0\,\mathrm{rad}\,\mathrm{m}^{-2}$, $\tilde{f}_{0}=0.84-100.93i$,
$\tilde{\phi}_{1}=62.4\,\mathrm{rad}\,\mathrm{m}^{-2}$, $\tilde{f}_{1}=-0.76+51.49i$.
and once again the ``dirty'' FDF and RM-CLEAN fail to recover the
Faraday depths, showing two main components at: $\tilde{\phi}_{0}=-6.4\,\mathrm{rad}\,\mathrm{m}^{-2}$,
and respectively $\tilde{\phi}_{1}=118.8\,\mathrm{rad}\,\mathrm{m}^{-2}$.
This is a typical example where the interference of pairs of RM components
conspire to place the peak response of the FDF outside of the RM range
between the two input RM components. In Figure 3 we consider the same
two components separated by $\phi_{1}-\phi_{0}\simeq0.25\delta\phi$,
$\phi_{0}=0\,\mathrm{rad}\,\mathrm{m}^{-2}$, $\phi_{1}=30.5\,\mathrm{rad}\,\mathrm{m}^{-2}$.
In this case the RM-MUSIC spectrum cannot separate the components anymore,
however it correctly shows only one maximum, situated between the two
input RM components, while RM-CLEAN suggests a single RM component outside the range of the input RMs.

From the above examples we see that RM-MUSIC performs better
than RM-CLEAN for high to moderate S/N cases. 
However, in low S/N situations the performance of RM-MUSIC 
deteriorates. It is quite difficult to quantify the S/N ratio exactly and to 
find a limiting S/N value, and more future evaluations of the method are 
required for various wavelength ranges and noise levels per channel. 
For the considered experiment layout our numerical simulations have shown 
that if the amplitude 
of the strongest RM component is becoming smaller than $|f_{0}|\simeq50$, and for a noise per channel 
 $\sigma=\sqrt{N}=12.247$, RM-MUSIC begins to behave more like RM-CLEAN, 
 and cannot separate correctly the input components. 
In Figure 4 we give such an example of two components with
the complex amplitudes $f_{0}=-30i$ and $f_{1}=15i$, Faraday
depths $\phi_{0}=0\,\mathrm{rad}\,\mathrm{m}^{-2}$ and $\phi_{1}=61\,\mathrm{rad}\,\mathrm{m}^{-2}$, 
separated by $\phi_{1}-\phi_{0}\simeq0.5\delta\phi$. One can see
that in this case, both RM-MUSIC and RM-CLEAN give similar results. We should also mention that the experiment layout used in simulations is very similar 
to the ASKAP's POSSUM 
survey \citep{g1}, where the expected final noise is $10$ $\mu$Jy/beam. This value suggests that for $N=150$ channels and a peak signal to noise threshold of $\sim4$, RM-MUSIC should work as indicated for sources with polarized flux higher than $0.5$ mJy/beam, which is rather bright but quite reasonable, since about $\sim57\%$ of the NVSS RM catalog sources fall into this category. 

\section{Conclusion}

We have discussed a novel deconvolution method for RM synthesis applications, 
based on an extension of the MUSIC super-resolution algorithm. Numerical results show that for high to moderate S/N cases 
RM-MUSIC outperforms the standard RM-CLEAN approach, being  able to recover Faraday depth values of closely spaced pairs of thin RM 
components, even in situations where the peak response of FDF is outside 
the range between the two input components. For low S/N cases, the RM-MUSIC and RM-CLEAN methods provide similar results. 
RM-MUSIC performs well for pairs of thin RM components, however it is not suitable when the input Faraday spectrum contains polarized 
emission at a continuous range of RM.

\label{lastpage}

\end{document}